\documentclass{iaus}
\usepackage{graphicx}

\newcommand{\msun}{M\ensuremath{_\odot}}        
\newcommand{\mdot}{\ensuremath{\dot M}}         
\newcommand{\ledd}{L\ensuremath{_{\rm Edd}} }   
\newcommand{\bh}{black hole}                    
\newcommand{\ao}{A0620--00}
\newcommand{\sgra}{Sgr A*}

\title[JD 11.~~ Jets at lowest mass accretion rate] 
{Jets at lowest mass accretion rates}

\author[Maitra et al.]   
{Dipankar Maitra$^1$,
Andrew Cantrell$^2$,
Sera Markoff$^3$,
Heino Falcke$^{4}$,
Jon Miller$^1$,
 \and Charles Bailyn$^2$}

\affiliation{
$^1$
Dept. of Astronomy, University of Michigan, \\ Ann Arbor, MI, USA 48109
\\ email: {\tt dmaitra,jonmm@umich.edu} 
\\[\affilskip]$^2$
Dept. of Astronomy, Yale University \\ New Haven, CT, USA 06511
\\email: {\tt andrew.cantrell,charles.bailyn@yale.edu}
\\[\affilskip]$^3$
Astronomical Institute ``Anton Pannekoek'', University of Amsterdam
\\ 1098 XH Amsterdam, The Netherlands
\\email: {\tt s.b.markoff@uva.nl}
\\[\affilskip]$^4$
Dept. of Astronomy, Radboud University
\\ 6500 GL Nijmegen, The Netherlands
\\email: {\tt H.Falcke@astro.ru.nl}
}

\pubyear{2010}
\volume{275}  
\pagerange{1--1}
\jname{Jets at all Scales}
\editors{G.E. Romero, R.A. Sunyaev \& T. Belloni, eds.}

\begin{document}

\maketitle

\begin{abstract} 
We present results of recent observations and theoretical modeling
of data from black holes accreting at very low luminosities (L/L$_{\rm
Edd}\lesssim10^{-8}$).
We discuss our newly developed time-dependent model for episodic
ejection of relativistic plasma within a jet framework, and a successful
application of this model to describe the origin of radio flares seen
in Sgr A*, the Galactic center black hole. Both the observed time lags
and size-frequency relationships are reproduced well by the model. 
%
We also discuss results from new Spitzer data of the stellar black
hole X-ray binary system $A0620$--$00$. Complemented by long term SMARTS
monitoring, these observations indicate that once the contribution from
the accretion disk and the donor star are properly included, the residual 
mid-IR spectral energy distribution of $A0620$--$00$ 
is quite flat and consistent with a non-thermal origin.
The results above suggest that a significant fraction of the observed
spectral energy distribution originating near black holes accreting at
low luminosities could result from a mildly relativistic outflow.
The fact that these outflows are seen in both stellar-mass black holes
as well as in supermassive black holes at the heart of AGNs strengthens
our expectation that accretion and jet physics scales with mass.
\end{abstract} 

\firstsection 
\section{Introduction} 
Collimated relativistic outflows or ``jets'' are observed to be very 
closely associated with compact accretors like black holes and
neutron stars where $GM/Rc^2$$\lesssim$$1$.  Such jets are known to emit across
a broad range of the electromagnetic spectrum, from radio
to X-rays, and perhaps even $\gamma$-rays.  Sometimes jets can be
imaged directly, but often their presence is inferred,
such as from flat to slightly inverted spectral energy distribution
(SED) in radio through IR (see, e.g., \cite[Markoff et al.
2001]{MFF2001}; \cite[Maitra et al. 2009a]{Maitraetal2009a};
\cite[Vila \& Romero 2010]{VilaRomero2010}), correlation between
fluxes (see, e.g., \cite[Gallo et al. 2003]{Galloetal2003};
\cite[Russell et al. 2010]{Russelletal2010}), and scaling relations
connecting radio and X-ray luminosities with the black hole mass
(\cite[Merloni et al.  2003]{Merlonietal2003}; \cite[Falcke et al
2004]{Falckeetal2004}; \cite[G{\"u}ltekin et al. 2009]{Gultekinetal2009}).


In X-ray binaries (XRB), compact steady jets appear to turn off
(on) as the source makes a transition from nonthermal to thermal
(thermal to nonthermal) X-ray state, at luminosities typically a
few percent of the Eddington luminosity.  See e.g.  \cite[Homan \&
Belloni (2005)]{HomanBelloni2005}, \cite[Remillard \& McClintock
(2006)]{RM2006} for details of X-ray states, and \cite[Fender
(2006)]{Fender2006} for a review on XRB jets.  However, there is
no clear consensus as to any lower luminosity limit for jets, or even 
about the mode of accretion at low
mass accretion rates (\mdot).  While it is generally agreed that the radiative
efficiency of the emitting plasma is low (e.g. \cite[Narayan \& Yi
1994]{NarayanYi1994}; \cite[Blandford \& Begelman
1999]{BlandfordBegelman1999}) at low \mdot,
it is not even clear whether the observed emission originates in
an inflow or an outflow.

Obviously instrumental limitations become important at the lowest
luminosities.  Nevertheless,
ever-improving technological advances are making it possible to
detect sources at luminosities as low as $10^{-9}$ L$_{\rm Edd}$,
a limit which is constantly decreasing.  
Results of these fascinating experiments point strongly to
the presence of outflows even at the
lowest luminosities, for stellar as well as supermassive black
holes.  Here we present a case study of two such sources: Sagittarius
A*, the supermassive black hole at the center of our galaxy, and
the stellar mass X-ray binary \ao, both of which suggest the
presence of a jet at luminosities $\lesssim 10^{-8}$ L$_{Edd}$.

\section{Sagittarius A*} 
\begin{figure}[!h]
\begin{center}
 \includegraphics[height=0.48\textwidth, angle=-90, clip=true, trim=10 20 00 10]{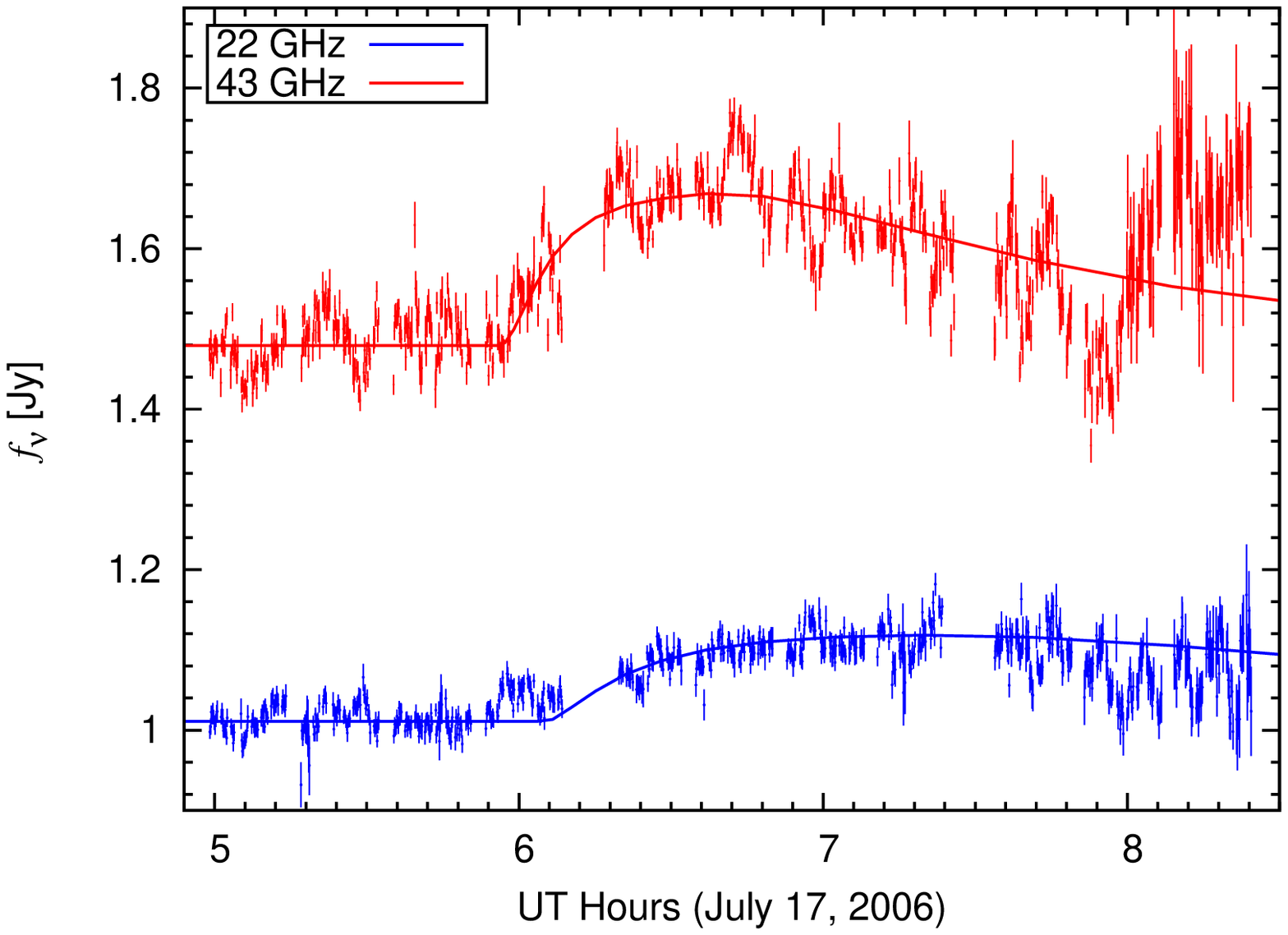}
 \includegraphics[height=0.48\textwidth, angle=-90, clip=true, trim=10 00 10 30]{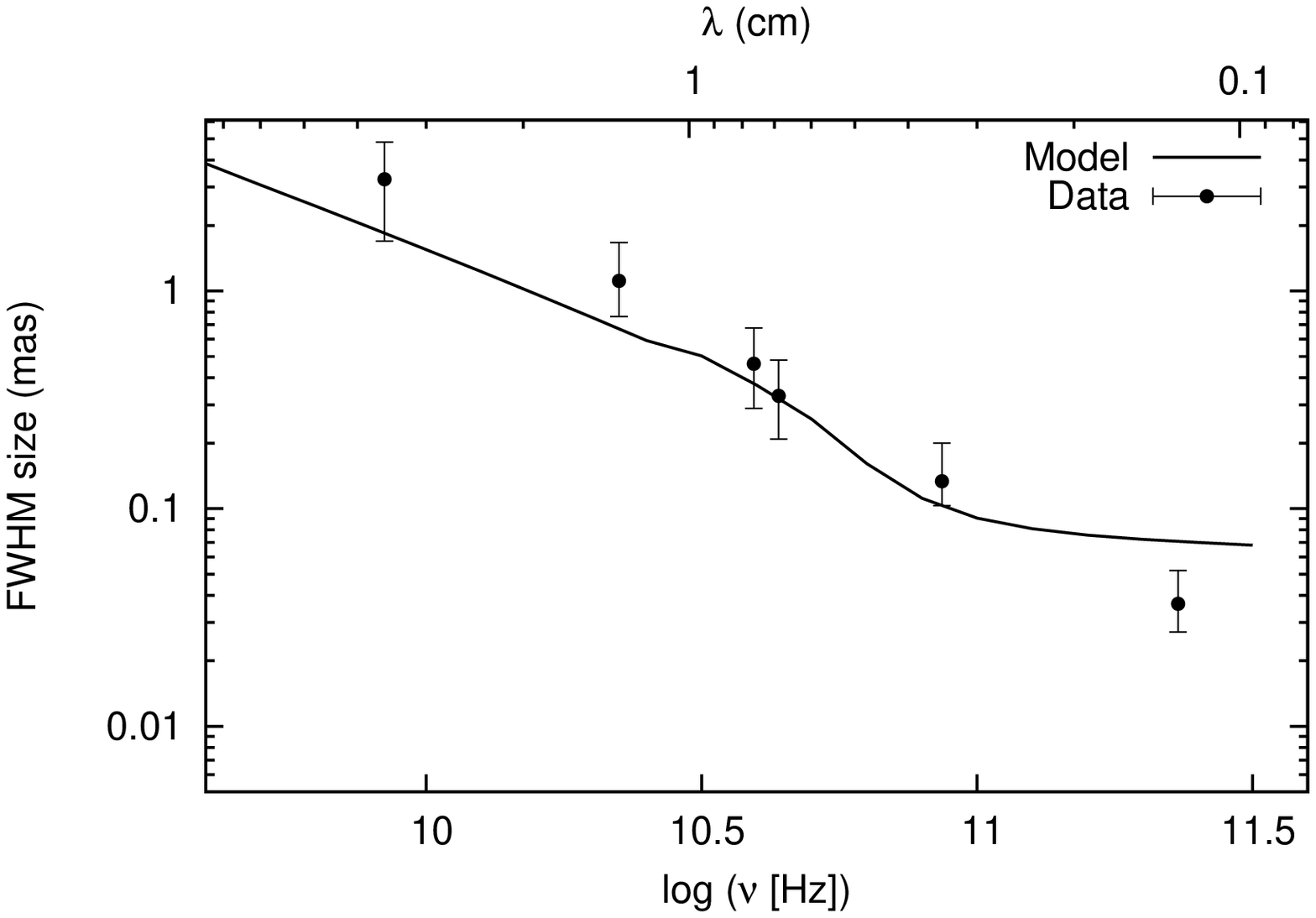}
 \caption{{\em Left:} Comparison of model with \sgra\ data for the flare on
 2006 July 17 at 43 and 22 GHz. The 43 GHz data with error bars from
 \cite[Yusef-Zadeh et al. (2008)]{YZ2008} are shown in red, and 22 GHz 
 data+model in blue. We model the flare that created a peak in the 43 GHz light 
 curve close to 6.5 hours UT.
 {\em Right:} Comparison of model-predicted frequency-size 
 relationship (solid line) with observations. The data are from 
 \cite[Bower et al. (2004)]{Boweretal2004}, 
 \cite[Shen et al. (2005)]{Shenetal2005}, and 
 \cite[Doeleman et al. (2008)]{Doelemanetal2008}. Figures taken from 
 \cite[Maitra et al. (2009b)]{Maitraetal2009b}.
}
   \label{f:sgra}
\end{center}
\end{figure}

Located $8.4\pm0.6$ kpc (\cite[Reid et al.
2009]{Reidetal2009}) away at the Galactic Center, Sagittarius A* 
(hereafter \sgra) is the nearest ``supermassive'' black
hole with a mass of $\sim4\times10^6$ \msun\ (\cite[Ghez et al.
2008]{Ghezetal2008}).  Despite its relative proximity, the source
of emission from \sgra\ is still unknown.  This is partly due to
the extremely high extinction in the direction of the Galactic
Center, and partly to the fact that with a
bolometric luminosity $\sim10^{-9}\ledd$, \sgra\ is extremely under-luminous 
compared to other active
galactic nuclei (AGN).
However, flares and
data from multiwavelength campaigns provide important clues about
the nature of emission from \sgra.  We have developed a time-dependent
jet model which for the first time allows comparing the model
predictions with radio flare data from \sgra.  Taking into account
relevant cooling mechanisms, we calculate the frequency-dependent
time lags and photosphere size and compare with recent observations
(see Fig.~\ref{f:sgra}). Details of the modeling is presented in
\cite[Maitra et al. (2009b)]{Maitraetal2009b}.

\section{A0620-00} 

Since its massive outburst in 1975, \ao\ has been in quiescence, with 
luminosities around $10^{-8}$ \ledd.  It has been detected in quiescence 
in X-rays, radio, IR, near-IR and optical using various space-based 
as well as ground-based instruments.

Despite easy detection in the optical and near-IR,
estimating the mass of the compact accretor has been quite a
challenge.  An accurate (dynamical) measurement of the accretor
mass requires correct knowledge of the orbital inclination.  Historically
attempts to measure the inclination from light curves have
given widely varying results with $38^{\circ}<i<75^{\circ}$.  However,
recently \cite[Cantrell et al. (2008,
2010)]{Cantrelletal2008,Cantrelletal2010} have analyzed $\sim30$
years of \ao\ optical and near-IR (OIR) data and showed that 
previous disagreements on the inclination of \ao\ were caused 
by (1) improper estimation of the disk flux (which can be 
significant even in quiescence), (2)
nightly variability in light curve shape, and (3) not
recognizing that even during quiescence, the OIR light curve 
shows three distinct optical states (named {\em passive}, {\em loop},
and {\em active}, and characterized by magnitude, color, and aperiodic
variability). Of these three states the passive state is the faintest,
shows consistent periodicity and is perhaps the closest to true
quiescence.  Consistent estimates of the binary parameters are
obtained when the above effects are properly accounted for, and the
best estimates of ($i$, M) are ($51.0^{\circ}\pm0.9^{\circ}$,
$6.6\pm0.25$ \msun). Accurate knowledge of the stellar contribution 
also allows determining any nonstellar residuals in the photometric
data.

\begin{figure}[h] 
\centering
 \includegraphics[height=0.48\textwidth, angle=90, clip=true, trim=35 050 80 20]{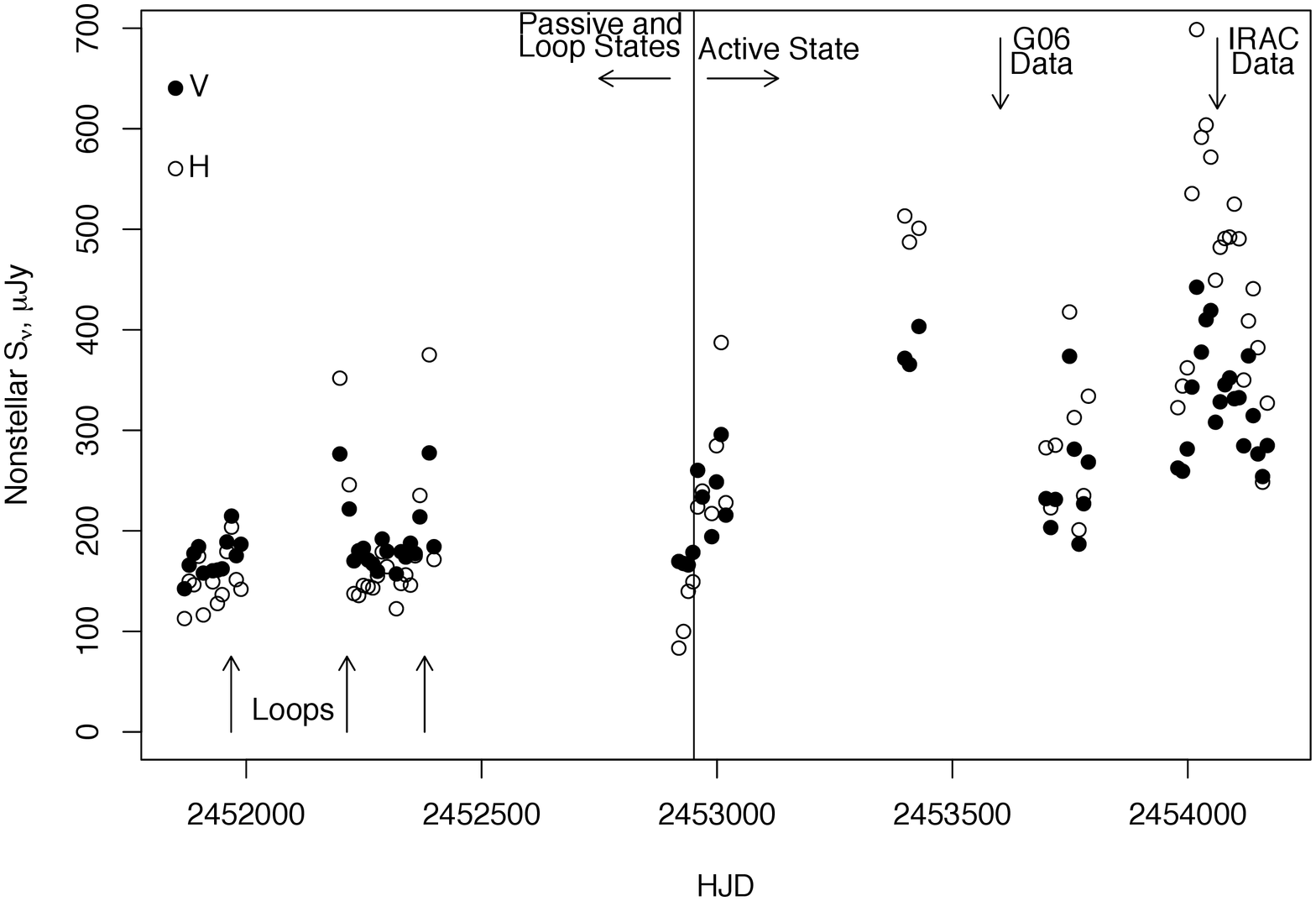}
 \includegraphics[height=0.48\textwidth, angle=00, clip=true, trim=20 180 25 20]{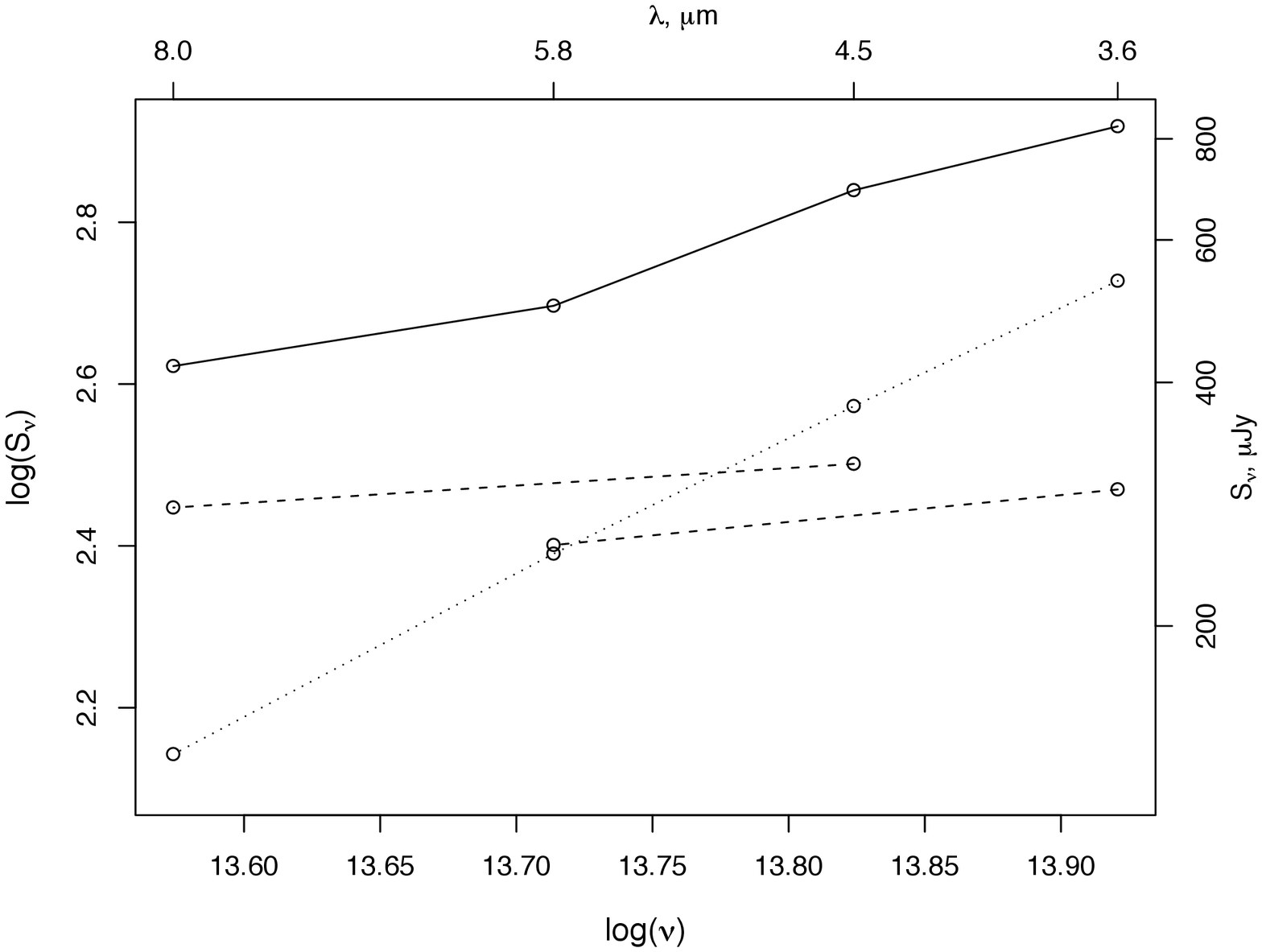}
 \caption{{\em Left:} Dereddened, nonstellar V- (filled points) and H-band 
 (open points) flux densities from SMARTS monitoring of \ao. OIR state 
 transitions as described in Cantrell et al 2010 are marked, as are the dates 
 of the {\em Spitzer}/IRAC data and the Gallo et al (2006) data. \ao\ is 
 consistently redder in the active state than in 
 the passive state, possibly due to a nonthermal contribution in the active 
 state.
 {\em Right:} The solid line shows the IRAC SED of \ao. Also 
 shown are the expected contributions from the donor + accretion disk 
 (dotted line), and the residual ``excess'' (observed SED minus the 
 donor+disk; dashed lines). The dashed lines connect the two pairs of
 bands which were obtained simultaneously. The excess has a relatively flat 
 spectrum with slope $\sim0.3$, suggesting that it could be at least partially 
 nonthermal.
}
   \label{f:ao}
\end{figure}

Here we present results of SMARTS long-term photometric monitoring
of \ao\ in OIR and a {\em Spitzer}/IRAC observation in 2006 November 
26.  Both the long-term monitoring as well as the {\em Spitzer}
snapshot suggest the presence of nonthermal emission, presumably
from a jet (see Fig.~\ref{f:ao}).

\section{Discussion and Conclusions} 
Recent multiwavelength data show that relativistic outflows or jets can be 
important at luminosities as low as $\sim$$10^{-9}\ledd$, and hence affect 
the evolution and energetics of compact accretors.  Here we have presented 
a case study of two low-luminosity accreting \bh\ 
sources, \sgra\ and \ao.  The two sources differ by five orders of magnitude in 
mass, but still the data point toward the presence of an outflow in both.

In particular, for \sgra\ we show that a time-dependent relativistic jet 
model can successfully reproduce:
\begin{itemize}
\item The quiescent broadband spectral energy
distribution of \sgra, 
\item The observed 22 and 43 GHz light curve
morphologies and time lags,
\item The frequency-size relationship. 
\end{itemize}
These results suggest that the observed radio emission from \sgra\
is most easily explained by a stratified, optically thick, mildly
relativistic jet.  Frequency-dependent measurements of time-lags
and intrinsic source size provide strong constraints on the bulk
motion of the jet plasma.

For \ao, radio, {\em Spitzer}, and SMARTS data obtained during 
quiescence suggest:
\begin{itemize}
\item Significant excess above the expected thermal emission from donor + 
accretion disk; spectral index of the excess -0.9$<$$\alpha$$<$-0.5 
($f_\nu$$\sim$$\nu^{\alpha}$), i.e. consistency with optically thin 
jet synchrotron emission in the NIR,
\item Nearly flat IR spectral index from {\em Spitzer} observations is 
suggestive of optically thick synchrotron emission in IR.
\end{itemize}

Techniques like Doppler tomograms of quiescent systems will lead
to better understanding of gas flow in the accretion disc of \ao\
and also that of the gas stream-disc impact point.
Future VLBI, submm and time-lag measurements over a broader range
of wavelengths will reveal greater details about ongoing physical
processes within a few gravitational radii for sources like Sgr A*
and M87.



\end{document}